\documentclass[12pt]{article}
\usepackage{graphics}
\begin{document}
\begin{center}
{\large Gamma Matrix Expansion of the Bethe-Salpeter Equation for Nucleon-Nucleon System }
\end{center}
\begin{center}
{Susumu Kinpara}
\end{center}
\begin{center}
{\it National Institute of Radiological Sciences \\ Chiba 263-8555, Japan}
\end{center}
\begin{abstract}
For the coefficients of the amplitude a set of simultaneous equations is derived in momentum space. 
By the auxiliary conditions they are equivalent to nonrelativistic equations
and suitable for the investigation of two-nucleon system. 
\end{abstract}
\section*{\normalsize{1 \quad Introduction}}
\hspace*{4.mm}
One of the interesting subjects in the field of the nuclear phenomena is the nuclear force and systematic investigations
of the nuclear structure and the reactions by the fundamental interaction.
It is traditional to calculate the observables for the nuclear many-particle system
in terms of the one-particle Green function within the field theoretical method. 
The meson-exchange picture and the resulting meson-exchange force are thus essential and a number of research
have revealed the spin-isospin properties of the finite density system.
Basically the shape of the potential between two nucleons is described by the screened Coulomb type (Yukawa potential type) 
and addition of the effect of the retardation does not affect the result of the calculation much 
in the Hartree-Fock approximation for the sd shell nuclei
since the difference in the value of the single particle energy between the occupied states is less than the mass of pion.
\\\hspace*{4.mm}
It is an interesting idea that the potential derived from the formulation of the one-particle Green function
for finite density system is applicable also to two-nucleon system in free space.  
Similar to one-particle case the formulation with the two-particle Green function is done
by equating the pole terms of the iteratively obtained integral equation. 
In the meson exchange model the nuclear force is originally expressed 
by the propagators of various mesons to understand the creation and annihilation processes between two nucleons.
The dependence of the potential on space-time requires to incorporate the effect of the retardation 
to avoid the divergence at the relative time $t=0$.
Then the form of the potential is thought to deviate from the Yukawa potential ($\sim {\rm exp} (-mr)/r$)
appropriate for calculations of the quantities on nuclear many-body system. 
Our purpose of this study is to derive the potential from the Bethe-Salpeter equation which describes 
two-nucleon system in the framework of the Lorentz invariance. 
\\\section*{\normalsize{2 \quad $\Gamma$ matrix expansion of the BS equation}}
\hspace*{4.mm}
The Bethe-Salpeter (BS) equation$\cite{Itzykson}$ is given 
to treat the bound state of nucleon-nucleon system and the form is
\\\begin{equation}
[\gamma\cdot(\frac{P}{2}+p)-M][\gamma\cdot(\frac{P}{2}-p)-M]\chi_P(p)=i\int \frac{d^4 p^\prime}{(2 \pi)^4}
V(p-p^\prime)\chi_P(p^\prime)
\end{equation}
in momentum space.
Here $\it P$, $\it p$ and $\it M$ are the total four-momentum, the relative four-momentum and the nucleon mass respectively.
The difference in the value of mass between proton and neutron is neglected. 
The BS amplitude $\chi_P(p)$ is a 4$\times$4 matrix as a function of the variable $p$ and also
contains a definite value of $P$ as denoted by the subscript.
\\\hspace*{4.mm}
It is the standard procedure to substitute the lowest-order diagram for the exact irreducible kernel 
which is known as the ladder approximation and we follow it in the present study.
The meson exchange model makes use of 
the isoscalar scalar sigma meson ($\sigma$), the isoscalar vector omega meson ($\omega$), 
the isovector pseudoscalar pion ($\pi$) and the isovector vector rho meson ($\rho$) as the source of the nuclear force.
The form of the interaction $V(q)$ is given in terms of the propagators as
\begin{eqnarray}
&&V(q)=\sum_{j}\frac{\Gamma_j}{m_j^2-q^2-i\epsilon},\;\;\;\;(j=\sigma,\omega,\pi,\rho)\\
&&\Gamma_\sigma=-g_\sigma^2,\\
&&\Gamma_\omega=g_\omega^2 \, \gamma \cdot \gamma,\\
&&\Gamma_\pi=-\,\vec{\tau_1}\cdot\vec{\tau_2}\,
(\frac{f}{m_\pi})^2 \, \gamma_{\rm 5} \gamma\cdot q \; \gamma_{\rm 5} \gamma\cdot q,\\
&&\Gamma_\rho=\,\vec{\tau_1}\cdot\vec{\tau_2}\,g_\rho^2 
\, (\gamma^\mu + ia \sigma^{\mu\nu} q_\nu) \, (\gamma_\mu - ia \sigma_{\mu\lambda}\, q^\lambda),
\end{eqnarray}
where $q \equiv p-p^\prime$.
\\\hspace*{4.mm}
Here $m_j$ ($j=\sigma,\omega,\pi,\rho$) are the masses of the mesons. The $g_\sigma$, $g_\omega$, $f$ and $g_\rho$
are the coupling constants of the $\sigma$, $\omega$ mesons, the pseudovector coupling constant of pion
and the vector coupling constant of $\rho$ meson respectively.
Concerning the $\rho$ meson exchange interaction the tensor coupling is also taken into account
and the mixing is given by $a\equiv\kappa/{\rm 2}M$ with the ratio $\kappa\equiv f_\rho/g_\rho$ 
where $f_\rho$ is the tensor coupling constant.
The $\vec \tau_i \; (i={\rm 1,2})$ are the usual isospin matrices for the $i$-th nucleon 
and hence $\vec{\tau}_1 \cdot \vec{\tau}_2 = {\rm -3}$ for the isospin singlet state ($T$=0) 
or $\vec{\tau}_1 \cdot \vec{\tau}_2 = {\rm 1}$ for the triplet state ($T$=1).
\\\hspace*{4.mm}
To proceed calculation Eq. (1) is transformed as 
\begin{eqnarray}
&&[\gamma\cdot(\frac{P}{2}+p)-M]\,\chi(p)_c\,[-\gamma\cdot(\frac{P}{2}-p)-M] \nonumber\\
&&
=\sum_j G_j \int \frac{d^4 p^\prime}{i(2 \pi)^4}
\frac{\Pi_j \,\chi(p^\prime)_c \,\Pi_j^\prime}{m_j^2-q^2-i\epsilon},\;\;\;\;(j=\sigma,\omega,\pi,\rho)
\end{eqnarray}
where 
$G_\sigma = g_\sigma^{\rm 2}$, 
$G_\omega = g_\omega^{\rm 2}$, 
$G_\pi = \vec{\tau_1}\cdot\vec{\tau_2}\,f^2 $ and
$G_\rho = \vec{\tau_1}\cdot\vec{\tau_2}\,g_\rho^{\rm 2}$
respectively.
The modified form of the BS amplitude is $\chi(p)_c \equiv \chi_P(p)\,C^{-1}$
utilizing the matrix to represent the charge conjugation $C$.
Thus the related quantities $(\Pi_j,\Pi_j^\prime)$ in Eq. (7) are given by 
\begin{eqnarray}
(\Pi_\sigma,\Pi_\sigma^\prime)&=&({\rm 1}, \,{\rm 1}),\\
(\Pi_\omega,\Pi_\omega^\prime)&=&(\gamma^\mu, \,\gamma_\mu),\\
(\Pi_\pi,\Pi_\pi^\prime)&=&(\gamma_{\rm 5} \gamma\cdot q \,m_\pi^{-1}, \,\gamma_{\rm 5} \gamma\cdot q \,m_\pi^{-1}),\\
(\Pi_\rho,\Pi_\rho^\prime)
&=&(\gamma^\mu+ia \sigma^{\mu\nu} q_\nu, \,\gamma_\mu-ia \sigma_{\mu\lambda} q^\lambda).
\end{eqnarray}
\hspace*{4.mm}
The modified BS amplitude $\chi(p)_c$ is expanded 
by means of a set of the Gamma matrices $\{\Gamma_\alpha\,\vert\,\alpha=1,\cdot\cdot\cdot,{\rm 16}\}$ as
\begin{eqnarray}
\chi(p)_c = \chi_S(p) + \gamma\cdot\chi_V(p) + \gamma_5\,\chi_P(p) + \gamma_5\,\gamma\cdot\chi_A(p)
-\sigma^\prime_{\mu\nu}\,\chi_T^{\mu\nu}(p),
\end{eqnarray}
where $\sigma^\prime_{\mu\nu}\equiv -i \gamma_5 \sigma_{\mu\nu}$
and hence $\chi_T^{\mu\nu}(p) = -\chi_T^{\nu\mu}(p)$.
The subscripts denote the scalar ($S$), the vector ($V$), the pseudoscalar ($P$), the axial-vector ($A$) and the tensor ($T$)
components respectively.
Similarly, the other invariants are
\begin{eqnarray}
\Pi_\omega\,\chi(p^\prime)_c \,\Pi_\omega^\prime = 4\,\chi_S -2\,\gamma\cdot\chi_V -4\,\gamma_5\,\chi_P
+2\,\gamma_5\,\gamma\cdot\chi_A,
\end{eqnarray}
\begin{eqnarray}
&&\Pi_\pi\,\chi(p^\prime)_c\,\Pi_\pi^\prime
=-\chi_S-\gamma\cdot\chi_V+\gamma_5\,\chi_P+\gamma_5\,\gamma\cdot\chi_A
-\sigma^\prime_{\mu\nu}\,\chi_T^{\mu\nu}\nonumber\\
&&+m_\pi^{-2}\,(q^2-m_\pi^2)(-\chi_S-\gamma\cdot\chi_V+\gamma_5\,\chi_P+\gamma_5\,\gamma\cdot\chi_A
-\sigma^\prime_{\mu\nu}\,\chi_T^{\mu\nu})\nonumber\\
&&+2\,m_\pi^{-2}\,q\cdot\chi_V\,\gamma\cdot q -2\,m_\pi^{-2}\,q\cdot\chi_A\,\gamma_5\gamma\cdot q \nonumber\\
&&-2\,m_\pi^{-2}\,\sigma^\prime_{\mu\nu}\,(q^\mu q_\lambda\,\chi_T^{\nu\lambda}-q^\nu q_\lambda\,\chi_T^{\mu\lambda}),
\end{eqnarray}
\begin{eqnarray}
&&\Pi_\rho\,\chi(p^\prime)_c\,\Pi_\rho^\prime
=4\,\chi_S-2\,\gamma\cdot\chi_V-4\,\gamma_5\,\chi_P+2\,\gamma_5\gamma\cdot\chi_A\nonumber\\
&&+a\,(-6\,q\cdot\chi_V+6\,\gamma\cdot q\,\chi_S
-4\,\gamma_5\gamma_\nu\, q_\mu \chi_T^{\mu\nu}-\sigma^{\prime\mu\nu}(q_\mu\chi_{A\nu}-q_\nu\chi_{A\mu}))\nonumber\\
&&+a^2(q^2-m_\rho^2)(3\,\chi_S+\gamma\cdot\chi_V
+3\,\gamma_5\,\chi_P+\gamma_5\gamma\cdot\chi_A+\sigma^\prime_{\mu\nu}\,\chi_T^{\mu\nu})\nonumber\\
&&+a^2m_\rho^2(3\,\chi_S+\gamma\cdot\chi_V
+3\,\gamma_5\,\chi_P+\gamma_5\gamma\cdot\chi_A+\sigma^\prime_{\mu\nu}\,\chi_T^{\mu\nu})\nonumber\\
&&-4\,a^2\,q\cdot\chi_V\,\gamma\cdot q -4\,a^2\,q\cdot\chi_A\,\gamma_5\gamma\cdot q.
\end{eqnarray}
Eqs. (12)$\sim$(15) construct the right hand side of Eq. (7).
In actual calculations the delta function parts in the $\pi$ and $\rho$ interactions are dropped as verified by 
differentiating the Feynman propagators in coordinate space.
\\\hspace*{4.mm}
The left hand side of Eq. (7) is expanded by Eq. (12) and the result is as follows
\begin{eqnarray}
&&[M-\gamma\cdot(\frac{P}{2}+p)]\,\chi(p)_c\,[M+\gamma\cdot(\frac{P}{2}-p)] \nonumber\\
&&
=h^S(p) + \gamma\cdot h^V(p) + \gamma_5\,h^P(p) + \gamma_5\,\gamma\cdot h^A(p)
-{\sigma^\prime}^{\mu\nu}\,h^T_{\mu\nu}(p),
\end{eqnarray}
\begin{eqnarray}
\qquad h^S(p)=(M^2-\eta+p^2)\,\chi_S(p)
-2Mp\cdot\chi_V(p)
-i\epsilon^{\mu\nu\rho\tau}P_\mu p_\nu \chi_{T\rho\tau}(p),
\end{eqnarray}
\begin{eqnarray}
&&h_\mu^V(p)=(M^2+\eta-p^2)\,\chi_{V \mu}(p)
+2p_\mu p\cdot\chi_V(p)
-\frac{P_\mu}{2}\,P\cdot\chi_V(p)\nonumber\\
&&-2M p_\mu\chi_S(p)
+i\epsilon_{\mu\nu\rho\tau}P^\nu p^\rho {\chi_A}^\tau(p)
-Mi\epsilon_{\mu\nu\rho\tau}P^\nu {\chi_T}^{\rho\tau}(p),
\end{eqnarray}
\begin{eqnarray}
\qquad h^P(p)=(M^2+\eta-p^2)\chi_P(p)
+M P\cdot\chi_A(p)
+2\,P^\mu p^\nu \chi_{T \mu\nu}(p),
\end{eqnarray}
\begin{eqnarray}
\qquad h_\mu^A(p)=(M^2-\eta+p^2)\chi_{A \mu}(p)
-2\,p_\mu\,p\cdot\chi_A(p)
+4 M\,p^\nu \chi_{T \mu\nu}(p)\nonumber\\
+\frac{P_\mu}{2}P\cdot\chi_A(p)
+M P_\mu\,\chi_P(p)
-i\epsilon_{\mu\nu\rho\tau} P^\nu p^\rho\,\chi_V^\tau (p),
\end{eqnarray}
\begin{eqnarray}
\quad && h^T_{\mu\nu}(p)=(M^2+\eta-p^2)\chi_{T \mu\nu}(p)
+2\,(p_\nu p^\sigma \chi_{T \mu\sigma}(p)-p_\mu p^\sigma \chi_{T \nu\sigma}(p))\nonumber\\&&
-\frac{P_\mu}{2}P^\sigma \chi_{T \sigma\nu}(p)
+\frac{P_\nu}{2}P^\sigma \chi_{T \sigma\mu}(p)\nonumber\\&&
-M p_\mu\,\chi_{A \nu}(p)
+M p_\nu\,\chi_{A \mu}(p)
+\,\frac{1}{2}(P_\mu\,p_\nu-P_\nu\,p_\mu)\chi_P(p)\nonumber\\&&
+\frac{M i}{2}\epsilon_{\mu\nu\rho\tau} P^\rho \,\chi_V^\tau(p)
-\frac{i}{2}\epsilon_{\mu\nu\rho\tau} P^\rho p^\tau\chi_S(p),
\end{eqnarray}
in which $\eta\equiv P^2/4$.
The simultaneous equations are obtained
by using the linearly independent relation ($\sum_\alpha c_\alpha \,\Gamma_\alpha = 0 \Rightarrow {}^\forall c_\alpha =0$)
for the set of the Gamma matrices in Eq. (7).
\\\section*{\normalsize{3 \quad The simultaneous equations of the BS amplitude}}
\hspace*{4.mm}
When $P_\mu = 0$ in the center of mass system the simultaneous equations become that of ref. $\cite{Nakanishi}$,
in which the system is divided into three groups $\chi_P$, ($\chi_V$, $\chi_S$) and ($\chi_A$, $\chi_T$) completely.   
On the other hand when $P_0 \neq 0$ the isolation of them does not continue  
and each component $\chi_\alpha$ is correlated with ones in the other groups.   
Still, it is shown that an isolation remains approximately if the auxiliary conditions are employed.
\\\hspace*{4.mm}
First of all we take up the vector equation ($c_V = 0$) and the vector components $\chi_{V \mu}(p)$.
Using the four-vector $\hat{P}_\mu \equiv P_\mu / \sqrt{P^2}$ 
it is assumed to be $\chi_{V \mu}(p) = \hat{P}_\mu \chi(p)$.
One advantage of it is that in the center of mass system ($P_i=0$) the terms about $\chi_{V \mu}(p)$
in $h_\mu^A(p)$ and $h^T_{\mu\nu}(p)$ are dropped exactly.
Besides there is the term $p\cdot\chi_V(p)$ in $h^S(p)$ 
and so $\chi_S(p)$ is required to be consistent with $\chi_{V \mu}(p)$.
\\\hspace*{4.mm}
Instead of the exact treatment of the simultaneous equations only for a narrow region of the possible solution 
in which the auxiliary relation 
\begin{eqnarray}
p\cdot\chi_V(p) = \hat{P}_0\,p_0 \chi(p) = 0
\end{eqnarray}
works they are investigated.
The solution of Eq. (22) is given by $\chi(p) \sim \delta (p_0)$ with the Dirac delta function.
The Fourier transform tells the point is such that $\partial \chi(x) / \partial t = 0$ in the coordinate space 
where $x^\mu \equiv x_1^\mu - x_2^\mu$ between two nucleons.
The equation may approximate to the exact one at the domain in which the relative time $t = 0$ 
since each potential in the right hand side of Eq. (7) gives $\partial V(x) / \partial t \vert_{t=0} = 0$
for $\sigma, \omega, \pi$ mesons and $\rho$ meson with $a \neq 0$ case
unless $V(x) \vert_{t=0} = 0$.
It is noted that for example when the tensor coupling of $\rho$ meson is turned on ($a\neq 0$) 
as seen from Eq. (15) the cross terms (the order $\sim O(a)$) may give rise to
$\partial V(x) / \partial t \vert_{t=0} \neq 0$ 
and therefore cause the violation of the auxiliary relation (Eq. (22)).
\\\hspace*{4.mm}
The zeroth vector component in the BS amplitude $\chi_P(p)$ ($\sim \gamma_0\,C\,\chi(p)$) is symmetric
under the transpose $(\gamma_0\,C)^T = +\gamma_0\,C$ while it gives the spin $S=0$.
The property of exchange between two identical fermions is unusual and 
then the zeroth vector component is not appropriate for the $S=0$ scattering wave to describe two nucleon system.
\\\hspace*{4.mm}
We pay attention to the pseudoscalar ($c_P$), the axial-vector ($c_A$) and the tensor ($c_T$) sectors now.
Following the vector component $\chi_{V \mu}(p)$ the axial-vector component 
is also assumed to be $\chi_{A \mu}(p) = \hat{P}_\mu \chi_a(p)$ 
and according to which the fifth term in $h_\mu^V(p)$ (Eq. (18)) is dropped.
While both $\chi_{A 0}(p)$ and the pseudoscalar component $\chi_P(p)$ are allowed 
to represent the spin singlet ($S=0$) state they have the potential different from each other.
\\\hspace*{4.mm}
The tensor component $\chi_{T \mu\nu}(p)$ is divided into two parts, that is, the polar components $\chi_{T 0 i}(p)$
and the axial components $\chi_{T i j}(p)$ ($i,j$=1,2,3).
We assume the axial components to be zero ($\chi_{T i j}(p) = 0$).
The triad of the polar components $\chi_{T 0 i}(p)$ constructs the spin triplet ($S=1$) state 
and symmetric under the exchange ($(\sigma^\prime_{0 i}\,C)^T = +\sigma^\prime_{0 i}\,C$).
The relative time dependence is neglected also in these components and then our interest is restricted to the region
$t = 0$ as well as the vector component $\chi_{V \mu}(p)$.
\\\hspace*{4.mm}
Retaining $\chi_a(p)$, $\chi_P(p)$ and $\chi_i(p) \equiv \chi_{T 0 i}(p)$ ($i$=1,2,3) 
and imposing the constraints mentioned above the equations $c_{P,A,T} = 0$ become
\begin{eqnarray}
&&(M^2-\eta-p^2)\chi_P(p) +P_0 \, F[p]\nonumber\\
&&=\sum_{j=\sigma,\omega,\pi,\rho} \lambda_j^P\,G_j \int \frac{d^4 p^\prime}{i(2 \pi)^4}
\frac{\chi_P(p^\prime)}{m_j^2-q^2-i\epsilon},
\end{eqnarray}
\begin{eqnarray}
&&(-M^2+\eta+p^2)\chi_a(p)
-2\,p_0^2\,\chi_a(p)
+2M F[p]\nonumber\\
&&=\sum_{j=\sigma,\omega,\pi,\rho} \lambda_j^A\,G_j \int \frac{d^4 p^\prime}{i(2 \pi)^4}
\frac{\chi_a(p^\prime)}{m_j^2-q^2-i\epsilon}\nonumber\\
&&-2m_\pi^{-2}G_\pi \int \frac{d^4 p^\prime}{i(2 \pi)^4}
\frac{q_0^2\,\chi_a(p^\prime)}{m_\pi^2-q^2-i\epsilon}\nonumber\\
&&-4a^2\,G_\rho\int \frac{d^4 p^\prime}{i(2 \pi)^4}
\frac{q_0^2\,\chi_a(p^\prime)}{m_\rho^2-q^2-i\epsilon}\nonumber\\
&&+\,4a\,G_\rho\int \frac{d^4 p^\prime}{i(2 \pi)^4}
\frac{q^i\,\chi_i(p^\prime)}{m_\rho^2-q^2-i\epsilon},
\end{eqnarray}
\begin{eqnarray}
&&(M^2-\eta-p^2)\chi_i(p) +2\,p_0^2 \,\chi_i(p) +p_i F[p]\nonumber\\
&&=\sum_{j=\sigma,\omega,\pi,\rho} \lambda_j^T\,G_j \int \frac{d^4 p^\prime}{i(2 \pi)^4}
\frac{\chi_i(p^\prime)}{m_j^2-q^2-i\epsilon}\nonumber\\
&&-2\,m_\pi^{-2}\,G_\pi \int \frac{d^4 p^\prime}{i(2 \pi)^4}
\frac{(q_0^2\,\delta^k_i+q_i q^k)\chi_k(p^\prime)}{m_\pi^2-q^2-i\epsilon}\nonumber\\
&&-\,a\,G_\rho \int \frac{d^4 p^\prime}{i(2 \pi)^4}
\frac{q_i\,\chi_a(p^\prime)}{m_\rho^2-q^2-i\epsilon},
\end{eqnarray}
\begin{eqnarray}
F[p] \equiv 2\,p^i \chi_i(p) +M \chi_a(p) +\frac{P_0}{2}\chi_P(p),
\end{eqnarray}
\begin{eqnarray}
p_0\,\chi_B(p) = 0 \qquad(B=P,\, a,\, i).
\end{eqnarray}
\\\hspace*{4.mm}
By virtue of Eq. (27) the constant solution at $t=0$ is thought to be a step in the right direction.
The left hand sides of Eqs. (23)$\sim$(25) are connected by $F[p]$.
It is responsible for the mixture of $L \pm 2$ components of the angular momentum in the tensor equation 
and then prompts us study the spheroidic properties of the tensor components.
In view of the fact on the quadrupole moment of deuteron it is possible to drop the term
$F[p]$ to make calculations of the elastic scattering tractable.
\\\hspace*{4.mm}
The coefficients $\lambda^{P,A,T}_{\sigma,\omega,\pi,\rho}$ in Eqs. (23)$\sim$(25) are shown in Table 1.
\begin{center}
   \begin{tabular}{|c|c c c c|}
   \multicolumn{1}{c}{ Table 1: Values of $\lambda^{P,A,T}_{\sigma,\omega,\pi,\rho}$ }\\
      \hline
equation & $\sigma$ & $\omega$ & $\pi$ & $\rho$ \\
     \hline
pseudoscalar ($P$) & 1 & $-4$ & 1 & $-4+3a^2m_\rho^2$ \\
     \hline
axial-vector ($A$) & 1 & 2 & 1 & $2+a^2m_\rho^2$ \\
     \hline
tensor ($T$) & 1 & 0 & 1 & $-a^2m_\rho^2$ \\
     \hline
   \end{tabular}
\end{center}
Consequently the strength of the interaction on the propagation of each meson
is determined by the value in Table 1 times the respective value of the coupling constant ($\lambda^{P,A,T}_j\,G_j$).
\\\hspace*{4.mm}
Some interesting features are seen in the meson exchange force.
The $\omega$ meson does not contribute to the tensor equation since the tensor coupling constant is usually set
nearly equal to zero $f_\omega \sim 0$.
It is noted that the derivative term ($\sim q_\mu q_\nu m_\omega^{-2}$) in the vector boson propagator 
is neglected since the mass is $5\sim6$ times as heavy as that of pion.
The contribution of the term to finite nuclei 
such as the closed shell nucleus ${}^{16}\rm O$ is only slightly because the difference of the single particle energy 
between two occupied states and the momentum transfer ($\vec{q}$) are much smaller than the mass of $\omega$ meson. 
\\\hspace*{4.mm}
Relating to sign of the interaction in the pseudoscalar equation $\omega$ meson 
is opposite to $\sigma$ meson making the isoscalar part cancel each other to some extent unlike 
the axial-vector equation.
Then the difference of the spin singlet component has an effect on the spin correlation parameter$\cite{Kinpara}$.
Except for the isospin and the coupling constant the interaction of pion 
is equal to that of $\sigma$ meson ($\lambda^{P,A,T}_\pi$=$\lambda^{P,A,T}_\sigma$) 
supporting $\sigma$ meson as the two-pion correlation.
An important feature of the pseudovector coupling of pion is the addition of the interactions
as seen in Eqs. (24) and (25).
Due to the quadratic four-momentum transfer ($q$) dependence 
the interactions are thought to work effectively not only in the asymptotic region
also in the high momentum transfer that is the short-range region.
In the $\rho$ meson interaction when the value of the vector-tensor ratio $\kappa$ is large 
the quadratically dependent terms may change the observables too.
The linearly dependent terms ($\sim a$) cause to connect the axial-vector and the tensor equations
by changing the angular momentum as $L \rightarrow L \pm 1$.
\\\section*{\normalsize{4 \quad Concluding remarks}}
\hspace*{4.mm}
In the present study we have obtained the simultaneous equations keeping the general properties of the BS equation.
As a consequence of the auxiliary conditions
they become the equivalent form to the eigen value equation.
The approximation about the relative time is suitable for the investigation of the BS amplitude in the space-like region
extremely far from the light cone where the meson exchange force is supposed to reach instantly.
Ultimately the nucleons are governed by the potential parts and in fact which may give the nuclear force
to describe two-body system.
Among various mesons pion is the lightest one and therefore has a particular effect.
In addition to the usual form of the propagator 
the pseudovector coupling interaction generates the inverse fourth power potential 
stemmed from the non-perturbative off-shell behavior of the BS equation.
The variation of the potential from the leading inverse square form is therefore large at the short-range
region, so that the cut-off is taken to avoid the divergence of the lowest-order term.
The procedure is also required to substitute the heavier mesons for one-pion exchange process to a degree.


\small
\begin{thebibliography}{99}
\bibitem{Itzykson}C. Itzykson and J.-B. Zuber, $\it{Quantum\, Field\, Theory}$, McGraw-Hill(1980).
\bibitem{Nakanishi}N. Nakanishi, Prog. Theor. Phys. Suppl. {\bf 43}, 1(1969).
\bibitem{Kinpara}S. Kinpara, arXiv:nucl-th/1604.00743.
\end{thebibliography}
\end{document}